\documentclass[useAMS,usenatbib]{mn2e}

\usepackage[dvips]{graphicx}
\usepackage{amssymb}
\usepackage{txfonts}

\input epsf.sty
\title[Masses of the components of the V1387 Aql/GRS1915+105 binary system]{On the masses
of the components of the V1387 Aql/GRS1915+105 binary system}
\author[J. Zi\'o{\l}kowski]{J. Zi\'o{\l}kowski \thanks{E-mail:
jz@camk.edu.pl}\\N. Copernicus Astronomical Center, ul. Bartycka
18, 00-716 Warsaw, Poland}
\begin{document}

\date{Accepted 0000 December 00. Received 0000 December 00; in original form 0000 December 00}

\pagerange{\pageref{firstpage}--\pageref{lastpage}} \pubyear{2014}

\maketitle

\label{firstpage}

\begin{abstract}

V1387 Aql (the optical companion to the microquasar GRS1915+105) is
a low mass giant. Such star consists of a degenerate, nearly
isothermal helium core and a hydrogen rich envelope. Both components
are separated by an hydrogen burning shell. The structure of such an
object is relatively simple and easy to model. Making use of the
observational values of the luminosity and of the radius of V1387
Aql, we determined the mass of this star as equal 0.28$\pm$0.02
M$_\odot$. This determination is relatively precise thanks to high
sensitivity of the luminosity of such structure to the mass of the
helium core and high sensitivity of its radius to the mass of the
envelope. The estimate does not depend on the knowledge of the
distance to the system (which is not precisely known). The main
source of the uncertainty of our estimate is uncertainty of the
effective temperature of V1387 Aql. When the effective temperature
will be known more accurately, the mass of V1387 Aql could be
determined even more precisely.

\end{abstract}


\section{Introduction}

GRS1915+105 is a low mass X-ray binary known also as one of the most
distinct Galactic microquasars (Mirabel \& Rodriguez 1994). The
system contains a black hole and a low mass K0-3 III optical
companion (Greiner et al. 2001) The companion which got a variable
star name V1387 Aql fills its Roche lobe and supplies the matter
accreted by the black hole. The first determination of the masses of
the components by Greiner et al. (2001) indicated that the system
contains one of the most massive galactic binary black holes with
the mass 14$\pm$4 M$_\odot$. This mass estimate was sensitive to the
distance to the system which was estimated from the kinematics of
the jets as 11--12 kpc (Mirabel \& Rodriguez 1994, Fender et al.
1999). After successful radio parallax measurement by Reid et al.
(2014) the distance was decreased to 8.6$^{+2.0}_{-1.6}$ kpc and the
revised mass of black hole to 12.4$^{+2.0}_{-1.8}$ M$_\odot$. This
new distance determination is consistent with independent estimate
of $\la$ 10 kpc by Zdziarski (2014) based on considering the jets
kinetic power.

Since the orbital period of the system is quite long (33.85 d), the
Roche lobe filling component must have large radius and cannot be a
main sequence star but rather an evolved giant (as indicated also by
the spectral luminosity class). The structure of such low mass giant
is relatively simple and easy to model. Making use of the
observational values of the luminosity and of the radius of V1387
Aql, we  will construct the model of this star and determine
relatively precisely (with the accuracy of $\sim$ 7\%) its mass.
This parameter was until now very poorly constrained (Fragos \&
McClintock (2015) gave the range (0.1--0.9 M$_\odot$).

\section{Observational parameters of V1387 Aql}

We will make use of very well known value of the orbital period of
the system ($P$ = 33.85$\pm$0.16 d, Steeghs et al. 2013), less well
known mass of the black hole component ($M_{\rm X} = 12.4
^{+2.0}_{-1.8}$ M$_\odot$, Reid et al. 2014) and the least well
known value of the effective temperature of V1387 Aql ($T_e =
4100$--5433 K, Fragos \& McClintock 2015). These parameters will be
used to estimate the radius and the luminosity of V1387 Aql.

From third Kepler law we have,

\begin{equation}
A/$R$_\odot = 4.208 (M/$M$_\odot (P/1 $d$)^2)^{1/3}
\end{equation}
where $A$ is the separation of the components, $M = M_{\rm X} +
M_{\rm opt}$ is the total mass of the binary system and $P$ is its
orbital period.

The relation between the radius of the optical component (which must
be equal to the radius of its Roche lobe) and the separation of the
components is (Paczy{\'n}ski 1971),
\begin{equation}
R_{\rm opt} = 0.462 A (M_{\rm opt}/M)^{1/3}
\end{equation}
where $M_{\rm opt}$ is the mass of the optical component.

From eqs. (1)$-$(2) we have,
\begin{equation}
R_{\rm opt}/$R$_\odot = 1.944 (P/1 $d$)^{2/3}(M_{\rm
opt}/$M$_\odot)^{1/3}
\end{equation}

From preliminary calculations we know that $M_{\rm opt}/$M$_\odot
\approx 0.28$ (the consecutive improvements of values of $R_{\rm
opt}$ and $M_{\rm opt}$ may be considered as quickly convergent
iterative process). Inserting this value and value for the orbital
period into eq. (3) we get $R_{\rm opt}/$R$_\odot = 13.31$. The
precision of this determination is $\sim$ 2\% (it is related to the
error of about 7\% in $M_{\rm opt}$).

Our radius estimate does not depend on the distance to the binary
system. However, having a relatively very precise radius estimate,
we may obtain a yet another estimate of the distance to the system.
We may use the observed angular diameter of the optical component.
The value of this diameter was estimated by Zdziarski et al. (2005)
from the $K$ band flux as equal $s$ = 0.0175--0.0220 mas. This value
together with our value of $R_{\rm opt}$ translates into the
distance 5.6--7.1 kpc. This range is marginally consistent with the
distance derived from the radio parallax by Reid et al. (2014).

To estimate the luminosity of V1387 Aql one needs to know its
effective temperature. It may be estimated from the spectral type of
the star: K0III--K3III (Fragos \& McClintock 2015). We will follow
the procedure of Fragos \& McClintock which was based on
calibrations tabulated by Gray (2008) and Cox (2000). This approach
leads to rather wide range of effective temperatures: 4100 to 5433 K
with the mean value of 4766 K (if one uses only Cox tables, the
range is much narrower: 4280 to 4660 K). The mean value of the
temperature together with the earlier determination of the radius
leads to the luminosity of 81.9 L$_\odot$ with the uncertainty range
of 45 to 138 L$_\odot$ (if one uses only Cox tables, the range is 53
to 75 L$_\odot$).

\section{The model of V1387 Aql}

The structure of low mass giant such as V1387 Aql is rather simple.
It contains a degenerate, nearly isothermal helium core surrounded
by a thin hydrogen burning shell. Above the shell there is a
hydrogen rich envelope. Its mass might be different for different
giants: it could be less than 10$^{-2}$ M$_\odot$ (as is the case
for V1387 Aql) but could be also several times greater than the mass
of the helium core. The luminosity generated by the hydrogen shell
(and so the luminosity of the giant) depends practically only on the
mass of the core. This relation (noted first by Paczy\'nski 1970) is
very well defined. This fact is demonstrated in Fig. 1.  One may
notice that the luminosity depends slightly also on the mass of the
envelope but this dependence is very weak: Fig. 1 shows that
increasing the mass of the envelope by a factor of 200, changes the
luminosity only by about 20 percent. I should comment on the fact
that the red curve was computed only for a relatively narrow range
of the helium core masses. This was due to the reasons of the
computational convenience: it was necessary to keep the very low
mass of the envelope while substantially changing the mass of the
underlying core. To summarize these considerations: the luminosity
of a giant translates directly into the mass of its helium core.

On the other hand, the radius of a giant depends mainly on its
luminosity (the effective temperature does not change very
significantly during this phase of the evolution) and so again
mainly on the mass of the core. The dependence on the mass of the
envelope is weak. This situation changes dramatically when the mass
of the envelope drops below about 10$^{-2}$ M$_\odot$: the radius
becomes a very sensitive function of the envelope mass. This is well
demonstrated in Fig. 2. If the star is in this evolutionary phase,
we can precisely determine the mass of the envelope from the radius
of the star.

We constructed models reproducing the present state of V1387 Aql
using the Warsaw evolutionary code described by Zi\'o{\l}kowski
(2005) . The standard Population I chemical composition of $X$ = 0.7
and $Z$ = 0.02 was adopted as the initial chemical composition. No
convective overshooting was taken into account. The results are
summarized in Figs. 1--3. From Fig. 1 we can read the present mass
of the helium core of V1387 Aql as 0.273$\pm$0.02 M$_\odot$. The
uncertainty of this determination was estimated from the upper
(blue) line in Fig. 1. Please, note that the dependance of the
luminosity on the core mass is so strong that in spite of the large
error of luminosity estimate (uncertainty by a factor of about
three) we still got quite precise (error of about 7\%) determination
of the core mass. In a similar way, Fig. 2 permits the determination
of the mass of the envelope of V1387 Aql as $\sim$
6.5$\times10^{-3}$ M$_\odot$. Together with the mass of the core it
gives the present mass of V1387 Aql as 0.28$\pm$0.02 M$_\odot$. Note
that this estimate is much more precise than the range 0.1--0.9
M$_\odot$ quoted by Fragos \& McClintock (2015).

\begin{figure}
\hbox{\epsfxsize=1\hsize\epsfbox{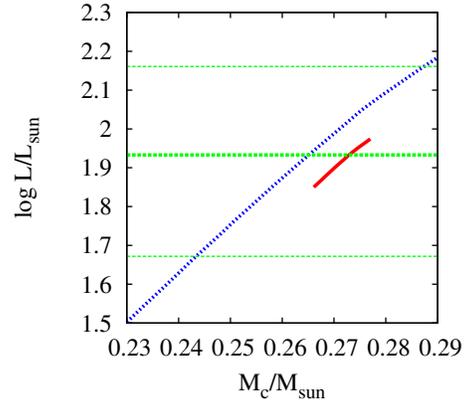}}
      \caption[h]{The core mass-luminosity relation for models
approximating the present evolutionary state of V1387 Aql. The lower
(red) curve describes models with the envelope mass equal about $6.5
\times10^{-3}$ M$_\odot$ (corresponding to the value for V1387 Aql).
The upper (blue) curve describes models with the envelope mass equal
about 1.25 M$_\odot$ (200 times larger). The thick horizontal
(green) line corresponds to the present luminosity of V1387 Aql and
thin horizontal lines correspond to the uncertainties of this
parameter.}
     \label{f1}
    \end{figure}

\begin{figure}
\hbox{\epsfxsize=1\hsize\epsfbox{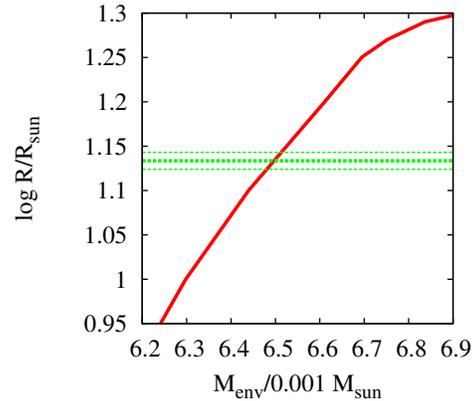}}
      \caption[h]{The envelope mass-stellar radius relation for models
approximating the present evolutionary state of V1387 Aql. The thick
horizontal (green) line corresponds to the present radius of V1387
Aql and thin horizontal lines correspond to the uncertainties of
this parameter.}
     \label{f2}
    \end{figure}

\begin{figure*}
\begingroup
   \def \A#1{\epsfxsize=0.48\hsize \epsfbox{#1}}
   \hbox to \hsize{\A{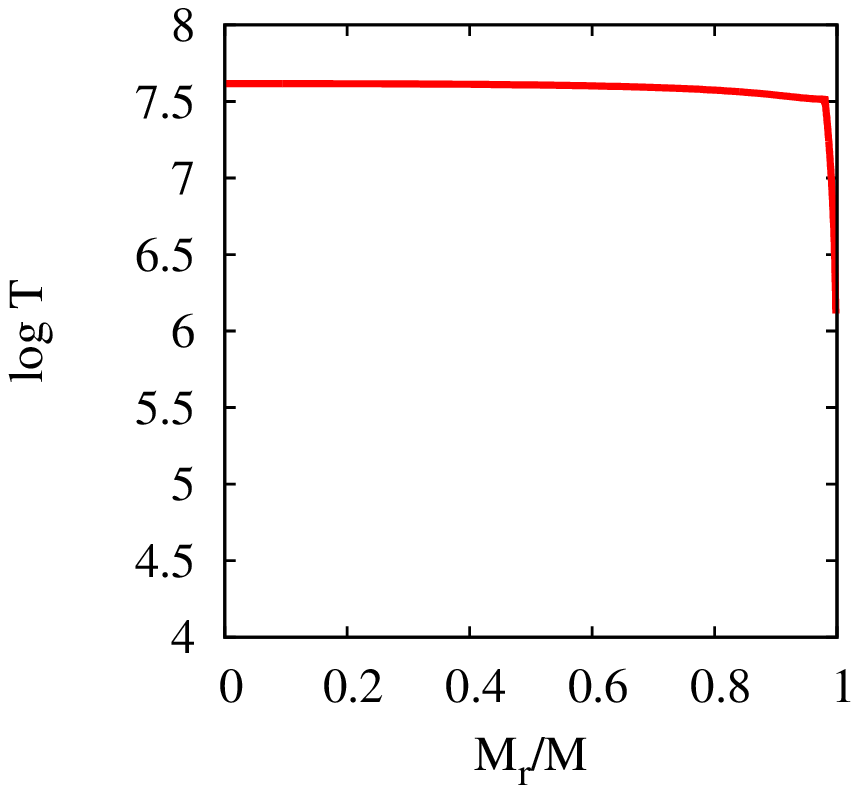}\hfil\A{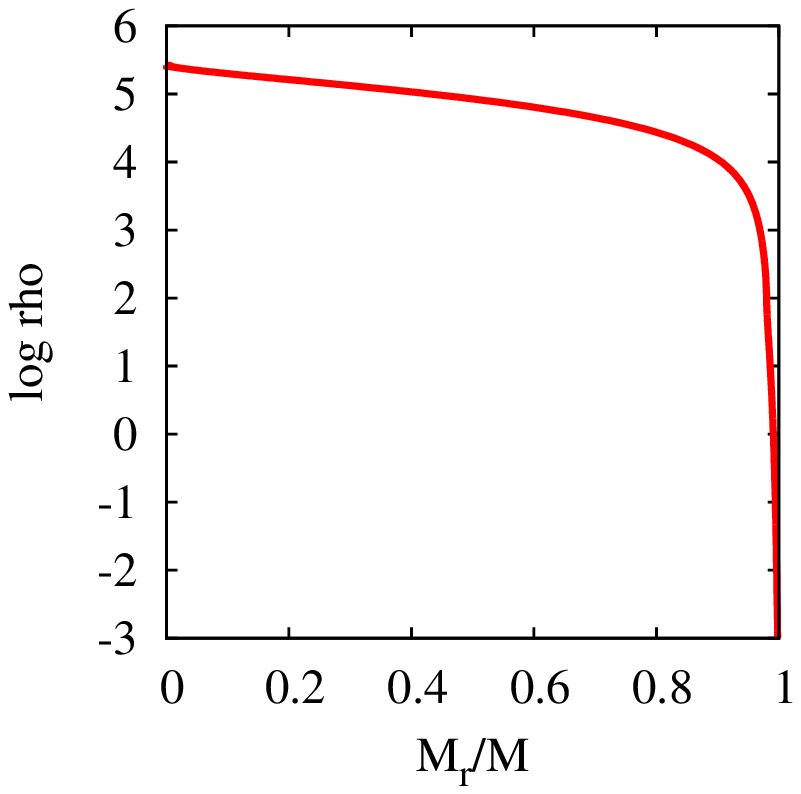}}
   \hbox to \hsize{\A{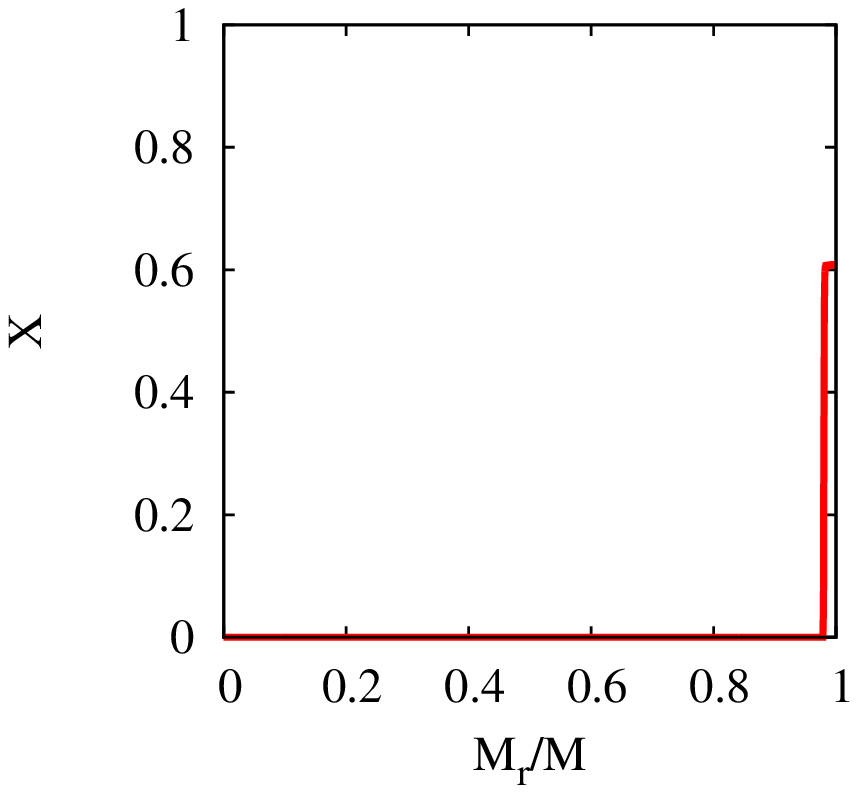}\hfil\A{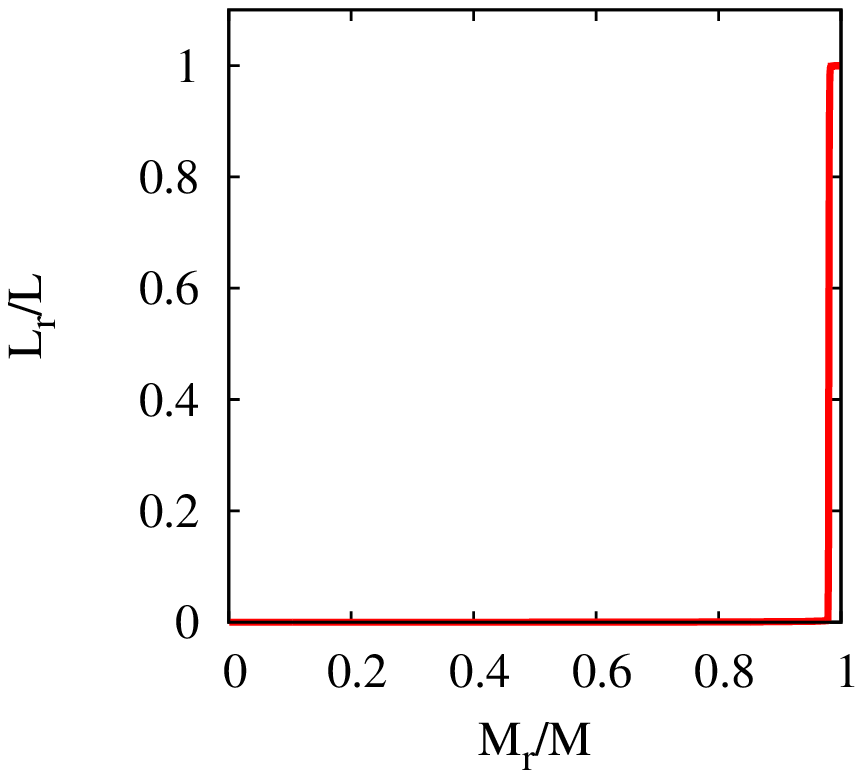}}
\endgroup
      \caption[h]{The internal structure of the model describing the
present evolutionary state of V1387 Aql: (a) the temperature
distribution, (b) the density distribution, (c) the hydrogen content
profile and (d) the distribution of the ratio of luminosity Lr
generated inside the radius r to the total luminosity L.}
     \label{f3}
    \end{figure*}

The internal structure of V1387 Aql is illustrated in Fig. 3. One
may see that indeed the core is nearly isothermal. I may add that
the degeneracy parameter of the electron gas $\psi = E_{\rm
Fermi}/$k$T$ is about 18.7 in the center of the star. It means that
the electron gas pressure and internal energy are by a factor of
about 7.5 larger than computed under the ideal gas approximation.

One comment that should be made concerns the luminosity of a star
losing mass through Roche lobe overflow. If the mass outflow is very
rapid, the surface luminosity of the star might be significantly
lower than in the case of no outflow. It is known that V1387 Aql is
losing mass to the black hole companion, occasionally at black hole
Eddington rate ($\sim 10^{-7}$ M$_\odot$/yr). However, this rate is
not high enough to decrease substantially the surface luminosity of
the star.

\section{The mass and the origin of the spin of black hole}

Unfortunately, the precise knowledge of the mass of the optical
component does not help to improve the accuracy of the black hole
mass determination. It would be possible if we knew the components
mass ratio precisely enough. Unfortunately, at present, the error of
the mass ratio determination is rather large: $q = M_{\rm
opt}/M_{\rm X}$ = 0.042$\pm$0.024 (Steeghs et al. 2013). This
determination is based on the rotational broadening of V1387 Aql
absorption lines. There is no hope that the precision of this
measurement will improve substantially in near future.

The present structure of V1387 Aql has no memory of its past
evolution. It could achieve the present state starting from
different initial configurations. Fragos \& McClintock (2015)
suggested that this initial configuration could be a star as massive
as 5 M$_\odot$. The large amount of the mass from the donor accreted
by black hole would then help to explain its large spin. It cannot
be excluded that V1387 Aql originated from the star of the initial
mass about 5 M$_\odot$. However, arriving at the present
configuration would require a rather fine tuning of its mass loss
history. It would have to be large during the initial phase of the
evolution to avoid the development of too large helium core.

\section*{Acknowledgements}

I would like to thank A. Zdziarski for a careful reading of the
manuscript and for helpful comments. This work was partially
supported by the Polish Ministry of Science and Higher Education
grant N203 581240 and by the Polish National Science Center project
2012/04/M/ST9/00780.

\label{lastpage}

\begin{thebibliography}{99}


\bibitem[\protect\citeauthoryear{Cox}{2000}]{b1}Cox, A.N., 2000, Allen's astrophysical quantities, 4th edn., ed. A.N. Cox
(New York: Allen's astrophysical quantities)
\bibitem[\protect\citeauthoryear{Fender et al.}{1999}]{b2}Fender, R.P., Garrington, S.T., McKay, D.J. et al. 1999,
MNRAS,304, 865
\bibitem[\protect\citeauthoryear{Fragos \& McClintock}{2015}]{b3}
Fragos, T., McClintock, J.E., 2015, ApJ, 800, 17
\bibitem[\protect\citeauthoryear{Gray}{2008}]{b4}Gray, D.F., 2008, The Observation and Analysis of Stellar Photospheres
\bibitem[\protect\citeauthoryear{Greiner et al.}{2001}]{b5}Greiner, J., Cuby, J.G., McCaughrean,
M.J., 2001, Nature, 414, 522
\bibitem[\protect\citeauthoryear{Mirabel \& Rodriguez}{1994}]{b6} Mirabel, I.F., Rodriguez,
L.F., 1994, Nature, 371, 46
\bibitem[\protect\citeauthoryear{Paczy\'nski}{1970}]{b7}Paczy\'nski B., 1970, AA, 20, 47
\bibitem[\protect\citeauthoryear{Paczy\'nski}{1971}]{b8}Paczy\'nski B., 1971, ARA\&A, 9, 183
\bibitem[\protect\citeauthoryear{Reid et al.}{2014}]{b9} Reid, M.J., McClintock, J.E., Steiner,
J.F. et al., 2014, ApJ, 796,2
\bibitem[\protect\citeauthoryear{Steeghs et al.}{2013}]{b10}Steeghs, D., McClintock, J.E., Parson,
S.G. et al., 2013, ApJ, 768, 185
\bibitem[\protect\citeauthoryear{Zdziarski}{2014}]{b11}Zdziarski, A., 2014, MNRAS, 444, 1113
\bibitem[\protect\citeauthoryear{Zdziarski et al.}{2005}]{b12}Zdziarski, A., Gierlinski, M., Rao, A.R., Vadawale, S.V., Miko{\l}ajewska, J., 2005, MNRAS, 360,
825
\bibitem[\protect\citeauthoryear{Zi\'o{\l}kowski}{2005}]{b13}
 Zi\'o{\l}kowski J., 2005, MNRAS, 358, 851


\end{thebibliography}
\end{document}